\documentclass[12pt, english]{article}

\usepackage{amsmath,amssymb,amsfonts}
\usepackage{graphicx}
\usepackage{dsfont}

\usepackage{fullpage}
\usepackage{bm}
\usepackage{siunitx}
\usepackage{caption}

\usepackage{algorithmic}
\usepackage{algorithm}

\floatname{algorithm}{Procedure}

\begin{document}

\title{Estimation of moisture content distribution in porous foam using microwave tomography with neural networks}

\author{Timo L\"ahivaara${}^a$, Rahul Yadav${}^a$, Guido Link${}^b$, and Marko Vauhkonen${}^a$\\
  ${}^a$Department of Applied Physics, University of\\ Eastern Finland, Finland\\
  ${}^b$Institute for Pulsed Power and Microwave Technology, Karlsruhe\\ Institut of Technology, Germany}

\maketitle

\subsection*{IEEE Copyright notice}

\textcopyright 20xx IEEE. Personal use of this material is
permitted. Permission from IEEE must be obtained for all other uses,
in any current or future media, including reprinting/republishing this
material for advertising or promotional purposes, creating new
collective works, for resale or redistribution to servers or lists, or
reuse of any copyrighted component of this work in other works.

\bigskip

\subsection*{Abstract}
 The use of microwave tomography (MWT) in an industrial drying process
 is demonstrated in this feasibility study with synthetic measurement
 data. The studied imaging modality is applied to estimate the
 moisture content distribution in a polymer foam during the microwave
 drying process. Such moisture information is crucial in developing
 control strategies for controlling the microwave power for selective
 heating. In practice, a reconstruction time less than one second is
 desired for the input response to the controller. Thus, to solve the
 estimation problem related to MWT, a neural network based approach is
 applied to fulfill the requirement for a real-time reconstruction. In
 this work, a database containing different moisture content
 distribution scenarios and corresponding electromagnetic wave
 responses are build and used to train the machine learning
 algorithm. The performance of the trained network is tested with two
 additional datasets.

\newpage
 
\section{Introduction}

Microwave drying is a process of heating a sample e.g. food products, wet wood, polymer foams using microwave sources \cite{Scherer, Makul}. The heating results in the removal of moisture from the samples due to penetration properties of microwave energy radiated from the microwave power sources and thereby producing a dry sample. While processing low-loss dielectric samples with non-uniform moisture distribution the problem of hot spot formation and thermal runaway have been observed as drying progresses. One solution to eliminate such issues is by selectively heating the sample by intelligent control of microwave power sources \cite{sun16}. The online control would adjust the power of the microwave source adaptively based on the knowledge of available moisture distribution inside the sample.  The infrared and temperature sensors, normally integrated with the microwave drying systems, are capable of giving information only on the surface which is not sufficient to provide efficient control of microwave sources.

Moisture measurement systems explicitly based on microwave technology have been utilized for the determination of moisture content in a sample during in-situ or ex-situ measurements \cite{Samir,Okamura2000,1634890,Trabelsi2009,8094000,6482167,7496964}. Such measurements are effective in giving quantitative information of the moisture content levels in the sample or limited to very local information, for example, on the sample surface but the actual moisture content distribution in the volume remains unknown. To provide information about the moisture content distribution inside the foam, a three-dimensional (3D) microwave tomography (MWT) system is proposed in this paper. Microwave tomography is a technique of estimating the material properties of an object from the measured data of scattered electromagnetic field around the object. The use case for MWT in industrial process imaging and its applications are detailed, for example, in \cite{15} and \cite{8016570}. 

For the studied microwave drying system employing a conveyor belt and large sample size, the speed by which the moisture distribution information will be available from the MWT is a challenge. Being a non-linear problem, image reconstruction in the MWT is a time consuming task since it requires solving the forward model multiple times. Popular choices of such iterative inversion algorithms applied in microwave tomography are, for example, Levenberg-Marquardt \cite{560338}, contrast source inversion \cite{10}, and subspace-based optimization method \cite{7422005}. However, due to the evaluation of the forward model multiple times, these methods may fail to provide estimates in real-time \cite{8741152, Sun:18} for online control. 

An attractive approach to fulfill the real-time estimation requirement is to use neural networks \cite{buduma17}. To train the network, we build a comprehensive synthetic database consisting of different moisture content distribution scenarios and the corresponding electromagnetic wave responses. The generation of such a database is computationally a demanding task but with the available computational resources, it can be simulated in a feasible time. Once the network is trained, it can be applied to recover the moisture content distribution in real-time.

Neural network approaches in MWT have been applied in various non-medical applications \cite{8476623, 8565987, 752198, 17, 1202957, 8709721} and an earlier implementation of neural networks in solving an inverse problem in electromagnetics was presented in \cite{312725} where the Hopfield neural network was exploited to solve Fredholm integral equation involving reconstruction of material properties of multilayered media. A three layer artificial neural network (ANN) system was implemented in \cite{728803} for determining the moisture content in wheat and in \cite{7951691} microwave reflection technique was used as a stimuli to the ANN built on an error backpropagation algorithm with momentum and adaptive learning techniques to predict the moisture content of commercially important biomass. Many other applications of neural network with microwave sensor technology are covered in \cite{24}. The moisture content distribution information inside the sample is estimated in \cite{LIM2003159, 25, 7801816} but they are limited to either small object sizes or low contrast objects. 

In this paper, the process of microwave drying is briefed in Section \ref{sec:drying}. Section \ref{sec:MWT} deals with the detailed modelling of the MWT in Comsol Multiphysics for collecting the synthetic scattered field measurements and a method to generate different moisture distributions in polymer foam is also explained. Section \ref{sec:neural} details the neural network based approach in the MWT and the results of the approach are shown in Section \ref{sec:Results}. Discussion and conclusions are given in Section \ref{sec:Discussion} and Section \ref{sec:Conc}.

\section{Drying system and an approach to integrate microwave tomography}\label{sec:drying}
  
Microwave drying is a complex phenomenon resulting from liquid evaporation via microwave heating. Microwave heating is directly associated with dielectric loss of material. A typical industrial microwave heating system, used for dielectric heating of materials in a continuous process, is based on high power microwave sources such as magnetrons, waveguide transmission lines, and microwave applicator where the material is processed on a conveyor belt. Electromagnetic energy is coupled into the microwave drying chamber or cavity through, for example, slotted waveguide or waveguide horn antennas. The power needed and the number of antennas to be used, in order to get maximum system efficiency, strongly depend on the system size, the amount of material to be processed and its dielectric behaviour during the process. However, it is observed that the system efficiency while processing low dielectric loss materials e.g., highly porous polymer foam, decreases with progress in drying as reported in \cite{guido}. 
  
\begin{figure}[!htb]
\centering
\includegraphics[width=0.6\textwidth]{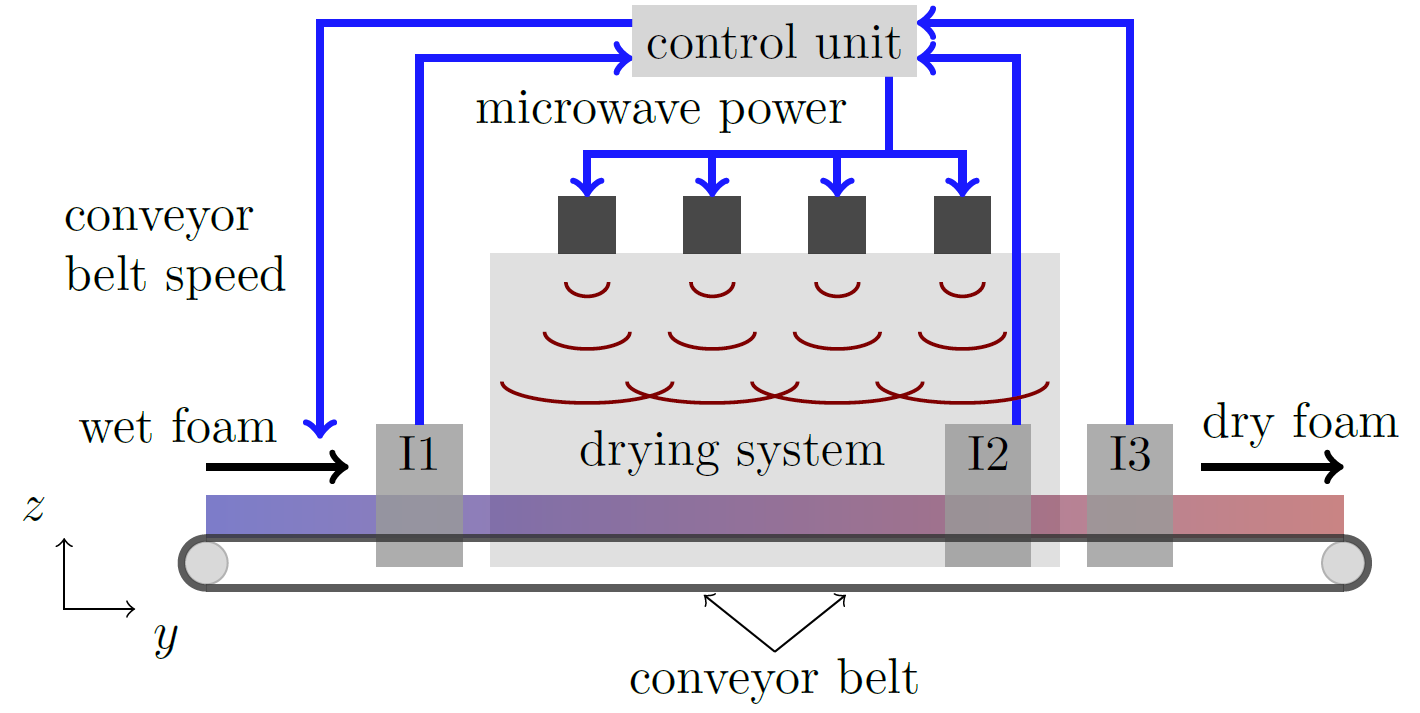}
\caption{Schematic showing cross-sectional view of important modules of a conventional microwave drying system.\label{fig:scemu}}
\end{figure}

A schematic picture of a microwave drying system is shown in Fig. \ref{fig:scemu}. The sample, in this case, a polymer foam with non-uniform moisture content distribution, enters the drying system from the left, propagates through the microwave applicator, and then exits the system from the right. During this process, the heterogeneity in the moisture content distribution can cause non-uniform heating. In addition, the non-uniform heating can lead to burning of the samples in the drying chamber due to hot-spot formation and thermal runaway condition. The ideal objective is to reduce the moisture content present in the polymer foam in minimum time with high system efficiency and no thermal runaway condition during the drying process. One way to tackle this issue is through selective heating by intelligent control of magnetron sources based on volumetric moisture distribution information. However, applying such a precise microwave control requires non-invasive in-situ measurement of the unknown distribution of moisture inside the material. The sensors already installed in the drying systems, like infrared and thermal cameras, shown as I2 block, are not sufficient to provide enough information for control purposes as their information is limited to the surface of the sample only. To make the moisture information available, an online microwave tomography sensor is proposed to be integrated outside of the system, shown as I1 and I3 blocks, for augmenting the control system of the microwave drying process. The control strategies based on the MWT output are not discussed in this paper.
 
\section{Model setup for the MWT}\label{sec:MWT}
 
Microwave tomography has two major steps i.e. to collect scattered electric field data around the object and to apply inversion algorithms on the scattered field data to get information on the object properties like dielectric distribution. In a microwave imaging system, the electric field data is acquired through an antenna array placed, in the near or far field, around the object. Some good calibration techniques \cite{6008622, 5617228, 7811280} need to be utilized for achieving better estimates from the reconstruction algorithm. Since our present focus is to first demonstrate the capabilities of MWT in drying systems, synthetic data is used as measurement data. The developed model includes dominant factors of real measurement scenarios like interferences or reflections and coupling between the antennas.

In a three-dimensional domain $\Omega$ with boundary $\partial \Omega$ characterized with relative permittivity $\epsilon_r$ and permeability $\mu_r$, the electric field $\vec{E}$ is governed by the vector-wave equation \cite{monk03, 5271011}
\begin{equation}
    \label{eq:maxwell}
    \nabla\times\left(\mu_r^{-1}\nabla \times \vec{E}\right)-k_0^2\epsilon_r \vec{E} =0, \quad \text{in}\quad\Omega,
\end{equation}
 where $k_0=2\pi/\lambda_0$ is the wavenumber and $\lambda_0$ is the wavelength in vacuum. The relative permittivity is expressed as a complex quantity $\epsilon_r=\epsilon_r^{\prime} - j\epsilon_r^{\prime\prime}$. In the following, the relative permeability $\mu_r$ is set to 1 for each material.

\begin{figure}[!htb]
\centering
\includegraphics[width=0.6\textwidth]{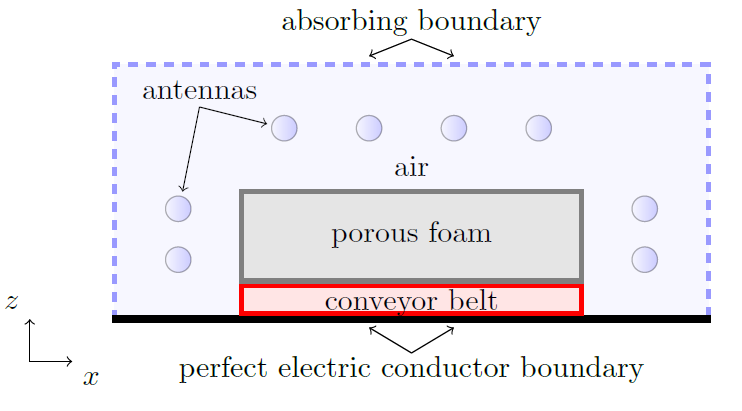}
\caption{\label{fig:geometry1}Cross-sectional view of the forward model for the free-space MWT system.}
\end{figure}

In this paper, a second order absorbing boundary condition is chosen for truncating the boundaries shown in Fig. \ref{fig:geometry1}. The bottom surface is modelled as a perfect electric conductor (PEC). More details on boundary conditions are available in \cite{5265547}.

\begin{figure*}[!htb]
\centering
\includegraphics[width=0.99\textwidth]{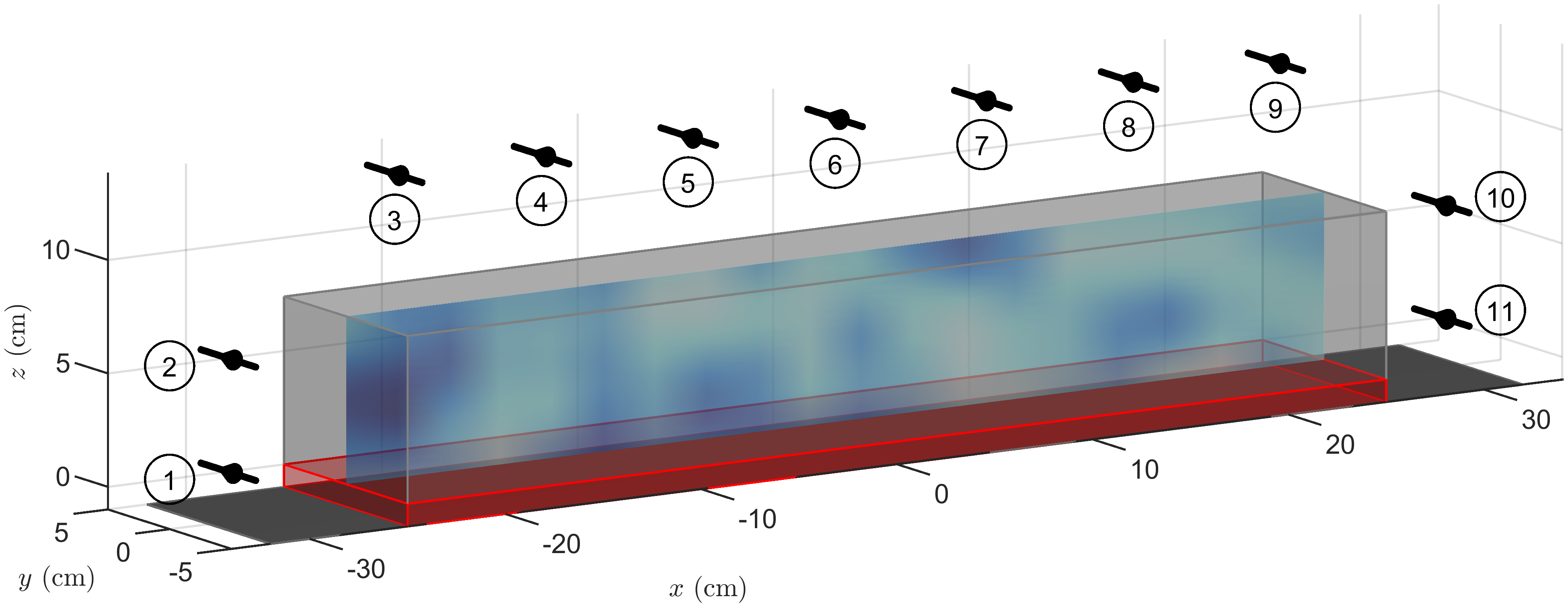}
\caption{\label{fig:geometry} Model used for benchmarking purposes. The light gray box is the porous foam and the red box the conveyor belt while dark gray surface denotes the PEC boundary. The antenna array is visualized with black cylinders and color surface inside the porous foam illustrates the potential moisture content distribution on the cross-section where $y=0$ cm.}
\end{figure*}

A finite element (FE) method based approach is resorted in this paper for solving the vector-wave equation. The FE model is constructed using a radio-frequency (RF) module in Comsol Multiphysics 5.3a \cite{comsol}. The computational domain consists of a porous foam $\Omega_f=[-25,\ 25]\times[-5,\ 5]\times[0,\ 7.4]$ cm and conveyor belt $\Omega_b=[-25,\ 25]\times[-5,\ 5]\times[-1,\ 0]$ cm, which are surrounded by air $\Omega_a=[-32,\ 32]\times[-5,\ 5]\times[-1,\ 14.4] \setminus (\Omega_f\bigcup\Omega_b)$ cm, see Fig. \ref{fig:geometry}. For the air subdomain, we set $\epsilon_r^{\prime}=1$ and $\epsilon_r^{\prime\prime} = 0$ and for the conveyor belt $\epsilon_r^{\prime}=1.1$ and $\epsilon_r^{\prime\prime} = 0$ whereas for the foam, only the relative permeability is assumed to be known accurately. The values of dielectric distribution are randomized through parametric models which are detailed in the following subsection. 

\begin{figure}[!htb]
\centering
\includegraphics[width=0.6\textwidth]{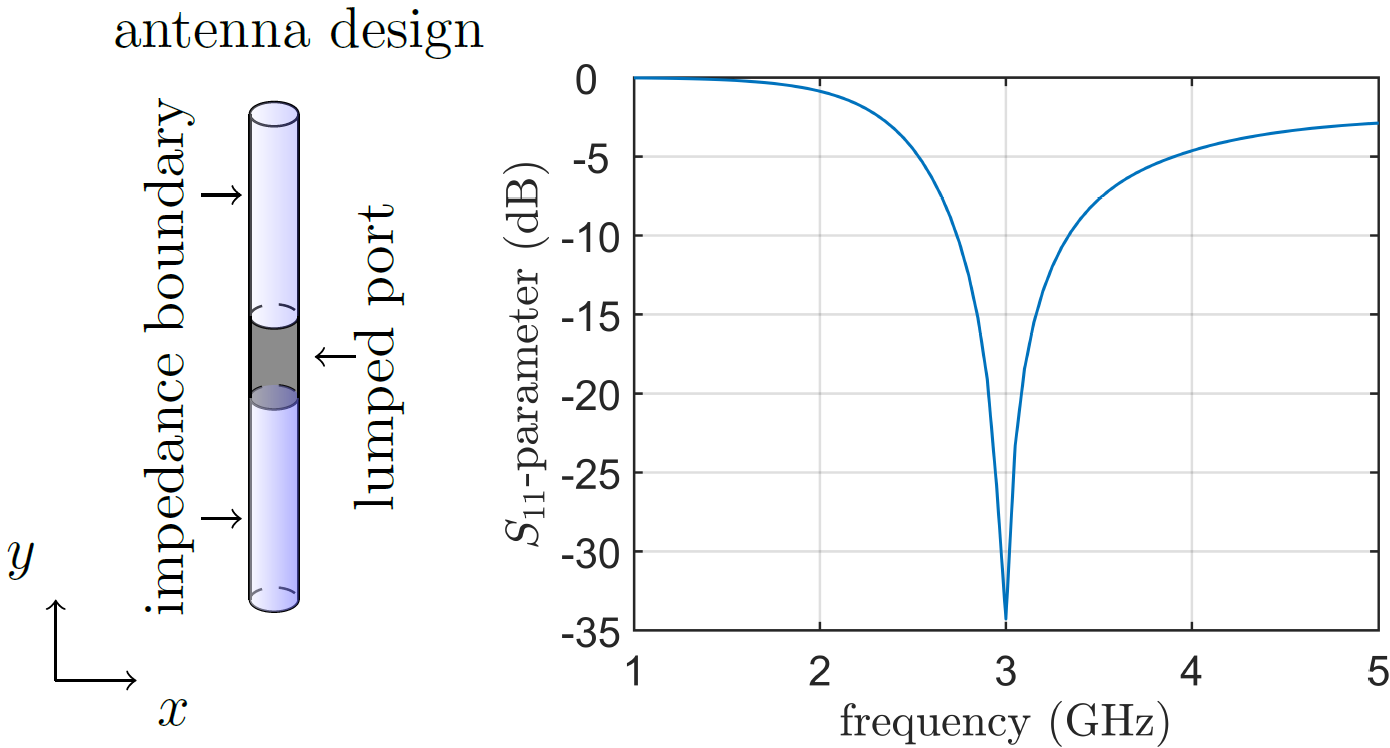}
\caption{\label{fig:antenna}Schematic of the dipole antenna with its boundary conditions, excitation and return loss response ($S_{11}$).}
\end{figure}

As a sensor setup, a dipole antenna with arm length = $18$ mm, radius = $1.03$ mm, and gap = $8.8$ mm with a resonant frequency of $3$ GHz is designed; 11 such dipole antennas are placed in a transceiver mode and are excited by a $75\si{\ohm}$ user defined lumped port. As the material for the sensor, we use copper with conductivity $\sigma =  5.998\times10^7$ S/m and $\epsilon_r=1$. In addition, they are modeled using an impedance boundary condition and polarized in the $y-$direction. The dipole antenna geometry and the return loss $(S_{11})$ are shown in Fig. \ref{fig:antenna}. The center point coordinates for the antennas are given in Table \ref{tab:antcoords}. 

In this work, the electric field measurements are simulated at $3\ {\rm GHz}$ frequency and are measured in terms of complex-valued $S$-parameter data. The measurement data is collected so that each antenna is separately in the source mode and it is assumed that it cannot receive data while acting as a source. Therefore, a total of $11\times10 =110$ measurements are performed. The scattered electric field is calculated by taking the difference between the measured complex-valued electric field with and without the foam in the forward model. 

In practice, the antenna setup is fixed and should measure the scattered field around the foam continuously in sync with the movement of the foam on the conveyor belt. Thus, the measurement acquisition system and the reconstruction scheme should be fast enough to estimate the moisture content distribution in the volume of the foam. A fast (and accurate) reconstruction will provide sufficient reaction time for the control system to achieve selective heating in the drying process. The focus of this work is on testing the reconstruction algorithm which can meet the real-time estimation requirement. Herein, the reconstruction scheme is evaluated only for a fixed cross-section (i.e. $x\times z$) of the foam.

\begin{table*}[ht]
  \centering
 \caption{Antenna center point coordinates.}\label{tab:antcoords}
\begin{tabular}{c||ccccccccccc}
 & 1 & 2 & 3 & 4 & 5 & 6 & 7 &8 & 9 & 10 & 11\\
\hline
$x_p$ (cm) & -31 & -31& -22.5& -15& -7.5&  0& 7.5& 15& 22.5& 31& 31\\
$y_p$ (cm) & 0 & 0& 0& 0& 0&  0& 0& 0& 0& 0& 0\\
$z_p$ (cm) & 1.2& 6.2& 13.4& 13.4& 13.4& 13.4& 13.4& 13.4& 13.4& 6.2& 1.2
\end{tabular}
\end{table*}

\subsection{Model parametrization for the porous foam}

A small volume of the foam is dielectric characterized at room temperature for different levels of moisture content $M_{{\rm meas}}$ using a waveguide transmission line technique with the foam covering the full waveguide cross-section. The real part of relative dielectric constant was found to be in the range of ${1.069}$ - ${1.795}$ and imaginary part varying between ${0.017}$ - ${0.098}$ for {\it dry basis} moisture content from 0 \% to 250 \%, respectively. From the curve fitting on the dielectric measurement data, the relationship between the {\it dry basis} moisture content $M_{{\rm meas}}$ and its corresponding dielectric value is obtained as
\begin{equation}
  \label{eq:param_true}
  \theta  =  \bar{a}_{\theta}\exp{(\bar{a}_{\theta}M_{{\rm meas}})},
\end{equation}
where $\theta = \{\epsilon_r^{\prime},\ \epsilon_r^{\prime\prime}\}$ denotes the material parameters. Numerical values for $\bar{a}_{\theta}$ and $\bar{b}_{\theta}$ are given in Table \ref{tab:parameters}.

\begin{table}[!htb]
  \centering
  \caption{Material model parameters.}\label{tab:parameters}
  \begin{tabular}{c||cc|cc}
  \hline
  & $\bar{a}_{\theta}$ & $\delta_{a_{\theta}}$ & $\bar{b}_{\theta}$ & $\delta_{b_{\theta}}$\\
  \hline
  $\epsilon_r^{\prime}$       & 1.0834 & 0.0151 & 0.0023 & 0.0001\\
  $\epsilon_r^{\prime\prime}$ & 0.0196 & 0.0013 & 0.0070 & 0.0004\\
      \end{tabular}
\end{table}

The procedure for generating the moisture content distributions $M$ and the corresponding dielectric permittivity values used in the neural network calculations is explained in the following. The moisture field variation inside the foam is assumed to be smooth. Here, such a random field is generated using a multivariate Gaussian distribution with anisotropic covariance structure, \cite{rasmussen}. The elements of the covariance $C$ can be calculated as
\begin{equation}
\label{eq:kernel}
  C_{ij} = \exp\left (  -\frac{\left \| x_i-x_j \right \|^2}{c_{x}^{2}}-\frac{\left \| y_i-y_j \right \|^2}{c_{y}^{2}}
   -\frac{\left \| z_i-z_j \right \|^2}{c_{z}^{2}}\right),
\end{equation}
where $i,j=1,\ldots,N_n$ and $c_x, c_y, c_z$ are the characteristic length components. $N_n$ denotes the number of locations. In practice, the characteristic lengths affects the correlation length of the moisture distribution in $x$, $y$, and $z$ directions and are randomized as $c_{x/y/z}\sim\mathcal{U}(3, 30)$ cm, where $\mathcal{U}$ denotes the uniform distribution.

The moisture content distribution $M$ can be expressed as
\begin{equation}
  \label{eq:moist}
  M  =  M^{\ast}\mathds{1} + \delta_{M}LZ,
\end{equation}
where $\mathds{1}$ is all-ones vector, $L$ is the lower triangular matrix of the Cholesky factorization of the covariance $C$,  $Z$ is a standard normal random vector, $M^{\ast}\sim \mathcal{U}(0, 200)\ \%,$  and $\delta_{M}\sim \mathcal{U}(0, 30)\ \%.$ In practice, $M^{\ast}$ and $\delta_{M}$ can be interpreted as the mean and standard deviation of the moisture content field, respectively. A pseudocode for generating a moisture sample is given below. 

\begin{algorithm}
\caption{Pseudocode for generating the moisture content distribution sample.  Note that a small diagonal component is added in matrix $C$ to ensure the positive definiteness.
}
\begin{algorithmic}[1]
\STATE Draw samples for $M^{\ast}$, $\delta_{M}$, $c_x, c_y, c_z$\smallskip
\STATE $C$ = ${\rm AnisotropicCovariance}(c_x, c_y, c_z, x, y, z)$
\STATE $L$ = ${\rm Cholesky}(C)$ 
\STATE $M$ = $M^\ast \, {\rm ones}(N_n) + \delta_M L \, {\rm randn}(N_n)$
\end{algorithmic}
\end{algorithm}

In simulating the electromagnetic wave responses for different moisture content realizations, the experimentally obtained mapping $M \to \{\epsilon_r^{\prime},\ \epsilon_r^{\prime\prime}\}$ is applied (see Eq. (\ref{eq:param_true})). In the current work, the relationship is replaced as
\begin{equation}
  \label{eq:param}
  \theta  =  a_{\theta}\exp{(b_{\theta}M)},
\end{equation}
where ${a}_{\theta}, {b}_{\theta}$ are random variables such as $a_{\theta}\sim\mathcal{U}(\bar{a}_{\theta}-\delta_{a_\theta}, \bar{a}_{\theta}+\delta_{a_\theta})$ and $b_{\theta}\sim\mathcal{U}(\bar{b}_{\theta}-\delta_{b_\theta}$, $\bar{b}_{\theta}+\delta_{b_\theta}$), where $\delta_{a_\theta}$ and $\delta_{b_\theta}$ define the bounds for the uniform distribution $\mathcal{U}$. Numerical values for $\delta_{a_\theta}$ and  $\delta_{b_\theta}$ are given in  Table \ref{tab:parameters}.

\begin{figure}[!htb]
\centering
\includegraphics[width=0.495\textwidth]{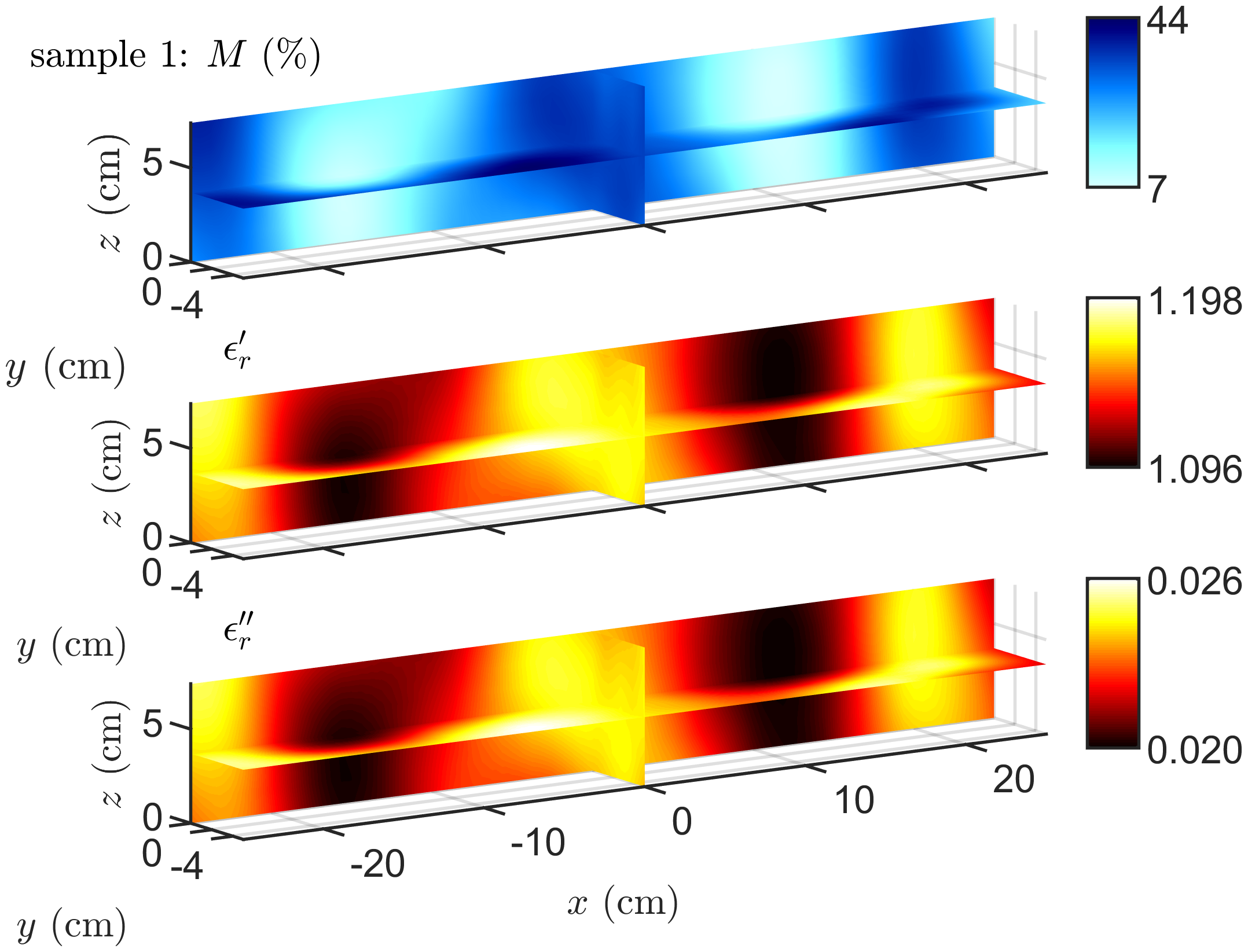}
\includegraphics[width=0.495\textwidth]{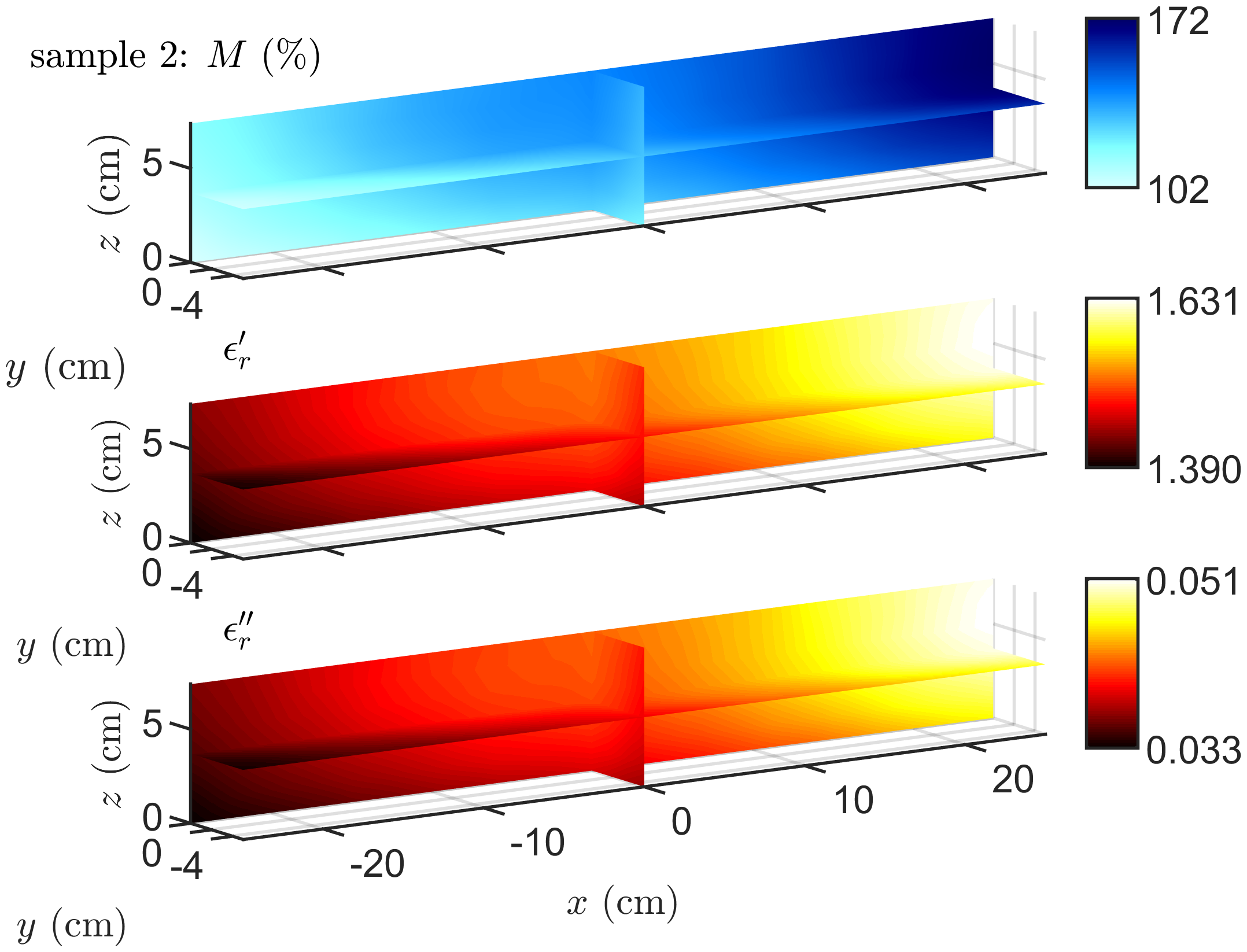}
\caption{\label{fig:samples} Moisture content distribution in {\it dry basis} weight \% (top), $\epsilon_r^{\prime}$ (middle), and $\epsilon_r^{\prime\prime}$ (bottom) for two samples. A foam sample with low (sample 1) and high (sample 2) water content. The characteristic length values $(c_x, \ c_y,\ c_z)$ for these cases are $(6.1,\, 4.8,\, 11.3)\ {\rm cm}$ (sample 1) and $(24.1,\, 14.2,\, 11.1)\ {\rm cm}$ (sample 2), respectively.}
\end{figure}

\begin{figure}[!htb]
\centering
\includegraphics[width=0.6\textwidth]{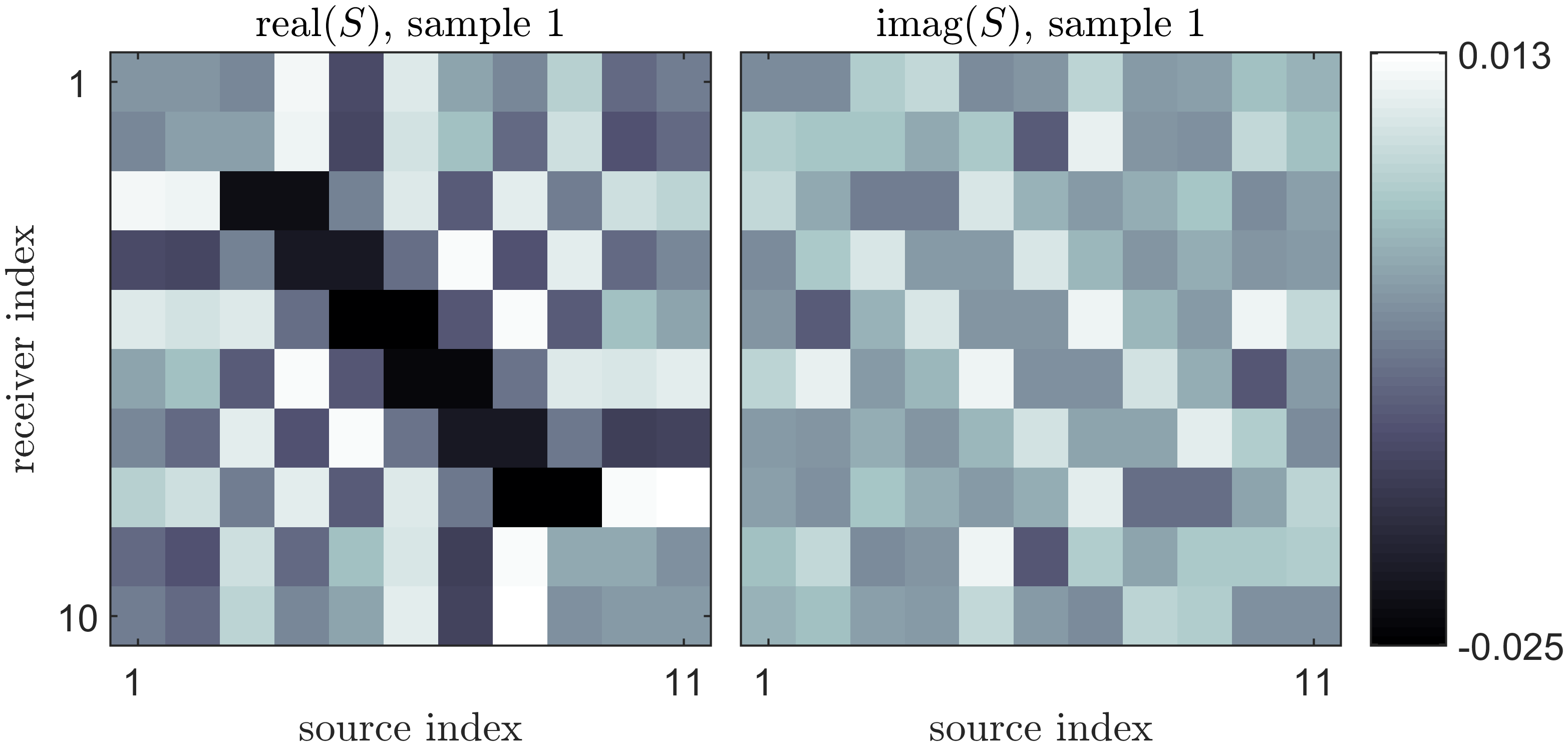}\\
\includegraphics[width=0.6\textwidth]{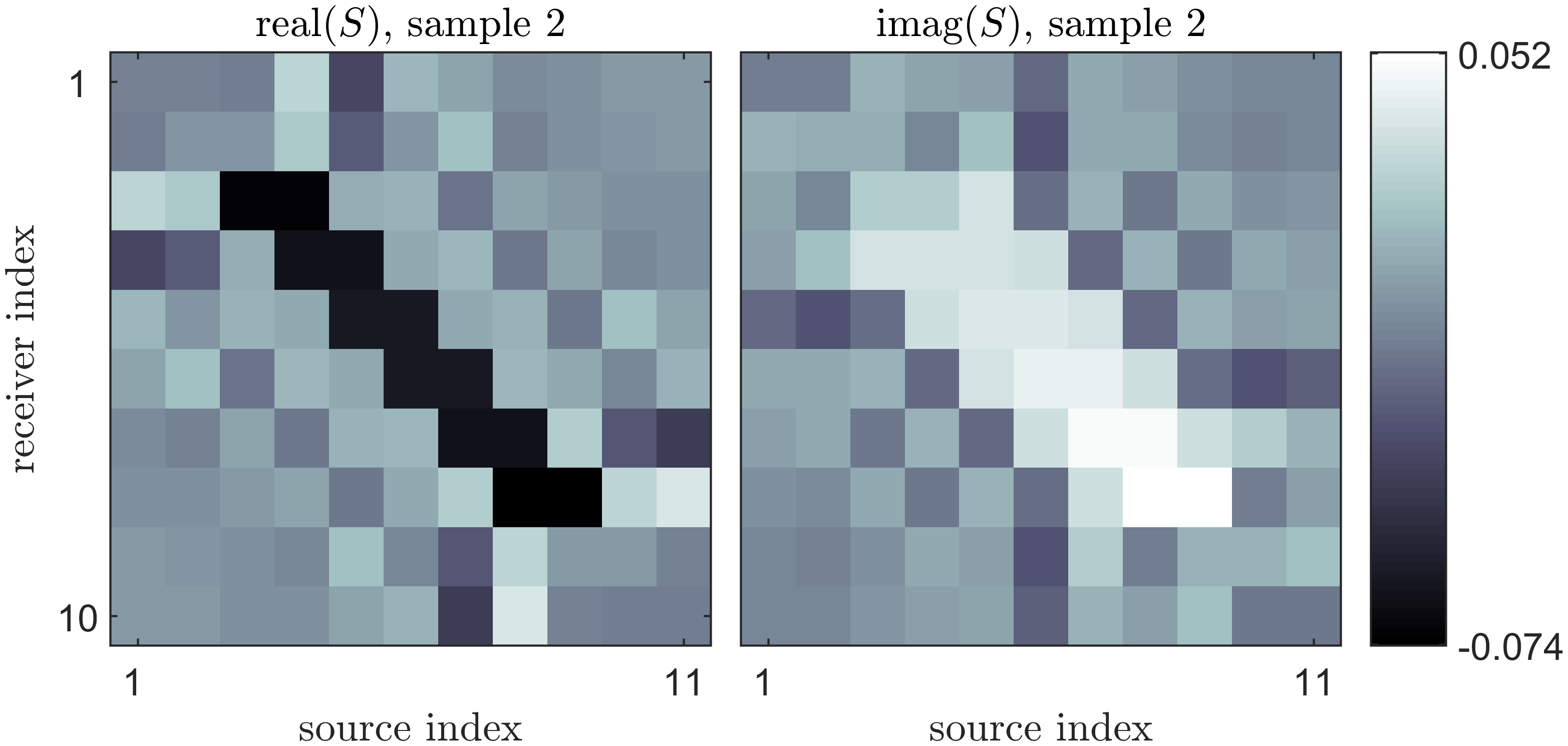}
\caption{\label{fig:data} Recorded $S$-parameter data corresponding to material samples shown in Fig. \ref{fig:samples}.}
\end{figure}

Two randomized material fields are shown in Fig. \ref{fig:samples}.  The figure shows the moisture content distribution and the corresponding physical parameters $\epsilon_r^{\prime}$ and $\epsilon_r^{\prime\prime}$. Sample shown on the top illustrates a situation where the characteristic length is greater in $z$-direction than in $x$- and $y$-directions, resulting in higher moisture variation in $z$-direction. Sample shown on bottom has greater characteristic length in $x$-direction than in $y$ and $z$, resulting in more moisture variation in $x$ direction. Figure \ref{fig:data} shows the recorded $S$-parameters for the material samples shown in Fig. \ref{fig:samples}. 

The computations were distributed to multiple computers operated by the research unit. As an example, the overall computational time for 220 $S$-parameter values shown in Fig. \ref{fig:data} (top) on a Linux computer equipped with Intel Xeon processor E5-1630 3.70GHz and 62GB random-access memory took around 10 minutes.

\section{Neural networks}\label{sec:neural}

In this work, we apply neural networks to estimate the moisture distribution of porous foam from microwave tomography scattered field data. The real and imaginary parts of the complex valued measurement data, i.e., $S$-parameter $X \in \mathbb{R}^{10\times11\times2}$ are given as an input to the neural network (real data is given on channel 1 and imaginary data on channel 2, respectively).  Since we are interested in monitoring a drying process of porous foam on a cross-section on $y$-axis, an adequate resolution of the moisture distribution field of around $x\times z = 2.4\times 2.5$ cm is chosen for the estimation. Thus, the moisture values are averaged in the $x$ and $z$ directions to form $21\times 3=$ 63 pixels.

\begin{figure*}[!htb]
\centering
\includegraphics[width=1.0\textwidth]{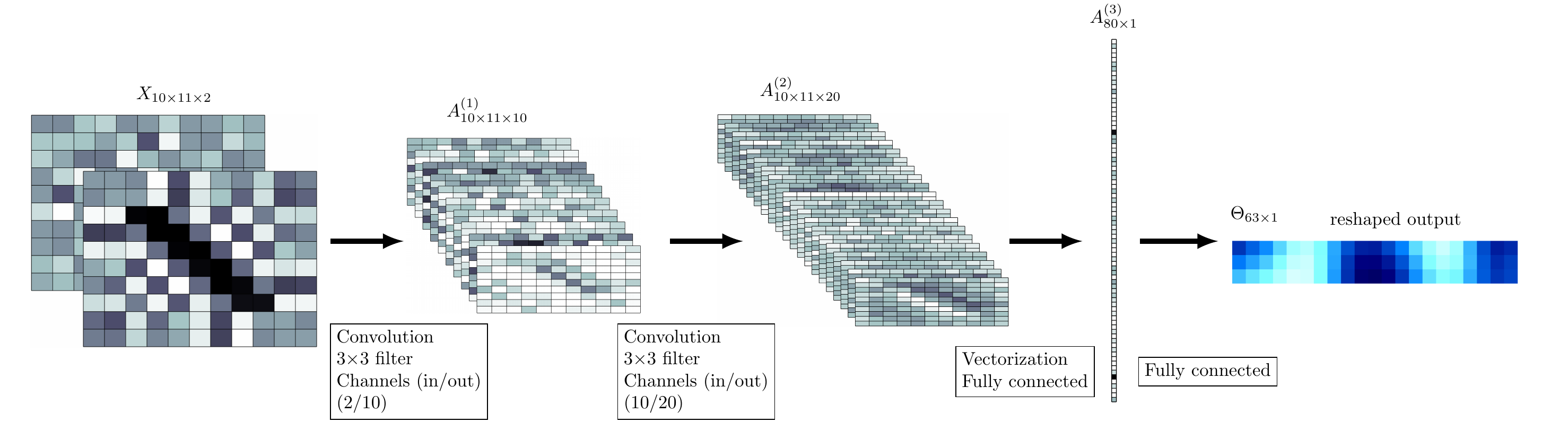}
\captionsetup{justification=centering}
\caption{\label{fig:tehtuuri} The neural network architecture used in this study.}
\end{figure*}

A neural network is trained to map from an input space $X \in \mathbb{R}^{10\times11\times2}$ to $\Theta \in \mathbb{R}^{63\times1}$ (vectorized moisture content distribution). The network architecture used in this work comprises two convolution layers and two fully connected layers.
The architecture used in this work has fewer convolution layers than commonly used for image classification, see e.g. \cite{krizhevsky}. In addition, instead of using a softmax layer for classification, the last layer is chosen to be a linear layer for regression. A similar type of network architecture has also been used by the authors in the context of porous material characterization \cite{lahivaara18, lahivaara19}. 
For the activation, we choose the non-linear Rectified Linear Unit (ReLU), expect on the output layer, where the linear layer is assumed. The network architecture is shown pictorially in Fig. \ref{fig:tehtuuri}.
The neural network $h_{b,w}$ is trained using a dataset comprising of moisture content distribution $\{\Theta_\ell\}$ and corresponding $S$-parameters $\{X_\ell\}$, $\ell=1,\ldots, N_{nn}$. $N_{nn}$ denotes the number of samples in the dataset. The generation of such a dataset is described below. In the training phase, the goal is to find biases $b$ and weights $w$ that minimize the discrepancy between $\{\Theta_\ell\}$ and the values predicted by the network $\{h_{w, b}(X_\ell) \}$.  In this work, we minimize the quadratic loss function to obtain the network parameters, biases, and weights of the network. The Adam optimizer \cite{Kingma2014} with the batch size of 200 samples was selected. For the network generation, a total of 2000 full training cycles in stochastic optimization (epochs) was run. A Python library TensorFlow \cite{tensorflow2015} was used as the computing interface. We set 1e-4 for the learning rate in the Adam optimizer, while all remaining parameters are left to default values used in Tensorflow. All neural network computations were performed on an Nvidia GeForce GTX 1080 graphics card.

\subsection{Training, validation, and test datasets}

For the training of the neural network, we generated a dataset comprising 15,000 samples, using computational grids that had $\sim$5 elements per wavelength. For each element, the quadratic basis function orders are set. The physical parameters $\epsilon_r^{\prime}$ and $\epsilon_r^{\prime\prime}$ for each sample were drawn using the framework discussed earlier. With the similar setup, 3000 samples were generated for validation. 

To train the network to tolerate the presence of the measurement noise, the training dataset responses were corrupted with simulated Gaussian noise. For the training database, we created five copies of each sample, leading to the total number of samples in the training set $N_{nn}=5\times 15000=75000$. Each of the responses for both datasets were corrupted as
\begin{equation}
\label{eq:noisemodel}  
X_{\ell}^\textrm{noised}=X_\ell+\alpha_{\max}\beta\epsilon,
\end{equation}
where $\alpha_{\max}$ is the maximum absolute value of the training dataset, the coefficient $\beta\sim\mathcal{U}(0.001,\ 0.022)$, and $\epsilon\sim\mathcal{N}(0,1)$ is sampled from normal distribution. In this work, the noise level is also expressed as a signal-to-noise ratio (SNR).  The noise parameter range $\beta\in [0.001,\ 0.022]$ leads to ${\rm SNR_{dB}} \in [15.7,\ 42.7]\, {\rm dB}$. The noise is added to validation dataset similarly as for the training samples. 

Furthermore, additional test samples were generated. The first test dataset, which comprises 3000 samples, was generated as the training data, except the computational grids were set to have $\sim$6 elements per wavelength. Different discretization was chosen to avoid the use of the same computational model to generate both training and test datasets. The use of the same model configuration could potentially lead to a situation in which severe modelling errors are ignored, yielding unrealistic impressions of the accuracy of the estimates as compared to actual performance with real measurement data \cite{kaipio07}. 

In this work, a spatially smooth variation of the moisture field inside the foam is assumed. We wanted to study the network's generalization capabilities by breaking the smoothness assumption. Therefore, we generated a second test dataset (a total of 100 samples) containing sharp material discontinuities inside the foam. As for the training data, we first used the model (\ref{eq:moist}) to generate moisture field and then updated some of the moisture field values as explained in the following. The number of rectangular cuboid shaped inclusions is randomized uniformly to be either 1 or 2. Each inclusion is assumed to infinite long in the $y$-direction, while the center point location $(x_{{\rm center}},\ z_{{\rm center}})$, cross-section size $(d_x,\ d_z)$, and uniform moisture content value $M_{{\rm inclusion}}$ are sampled as
\begin{eqnarray*}
\left(x_{{\rm center}},\ z_{{\rm center}}\right) &\sim& \left(\mathcal{U}(-25,\ 25),\ \mathcal{U}(0,\ 7.4)\right)\ {\rm cm},\\
\left(d_x,\ d_z\right) &\sim& \left(\mathcal{U}(10,\ 20),\ \mathcal{U}(3,\ 6)\right)\ {\rm cm},\\ 
M_{{\rm inclusion}} &\sim&  \mathcal{U}(0,\ 200)\ \%.
\end{eqnarray*}

One must note that for both test datasets, measurement noise was added more systematically to study the performance with different noise levels.

\section{Results}\label{sec:Results}

This section gives results to evaluate the performance of the proposed neural network based estimation scheme. We applied the trained neural network to predict the cross-section of the 3D moisture field of the test datasets. The results are shown for three special cases (Sections \ref{sec:low}, \ref{sec:uni}, and \ref{sec:gen}) and network's sensitivity to measurement noise is also evaluated (Section \ref{sec:noise}).

\subsection{Sample with low, moderate, and high moisture content}\label{sec:low}

\begin{figure*}[!htb]
\centering
\includegraphics[width=0.99\textwidth]{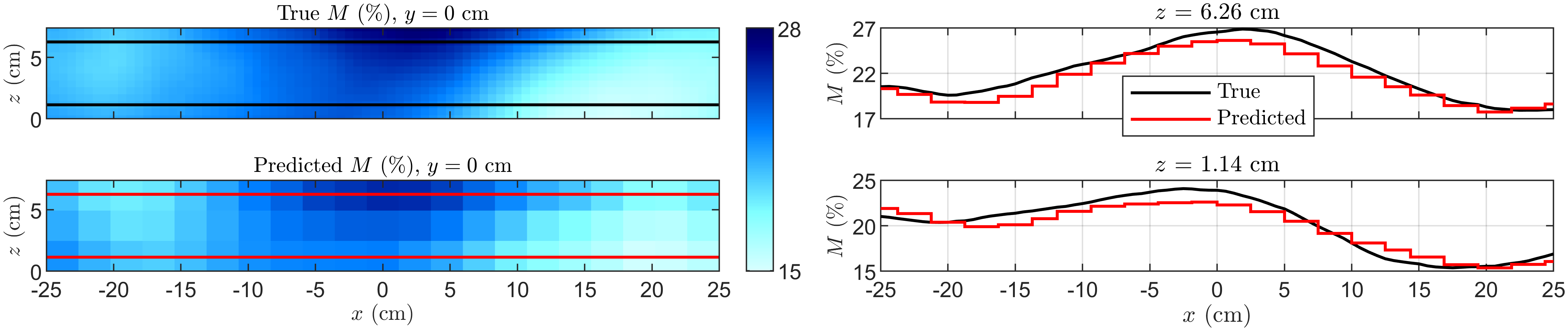}
\includegraphics[width=0.99\textwidth]{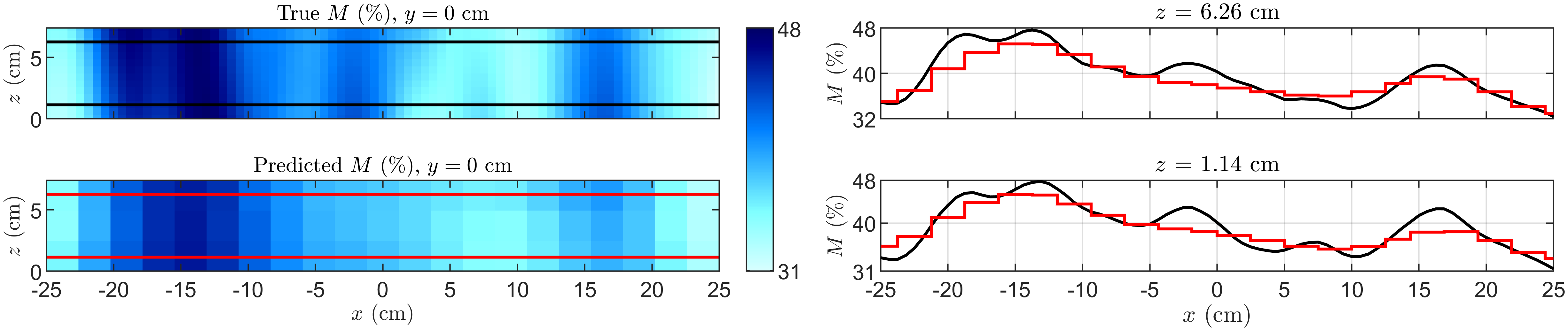}
\includegraphics[width=0.99\textwidth]{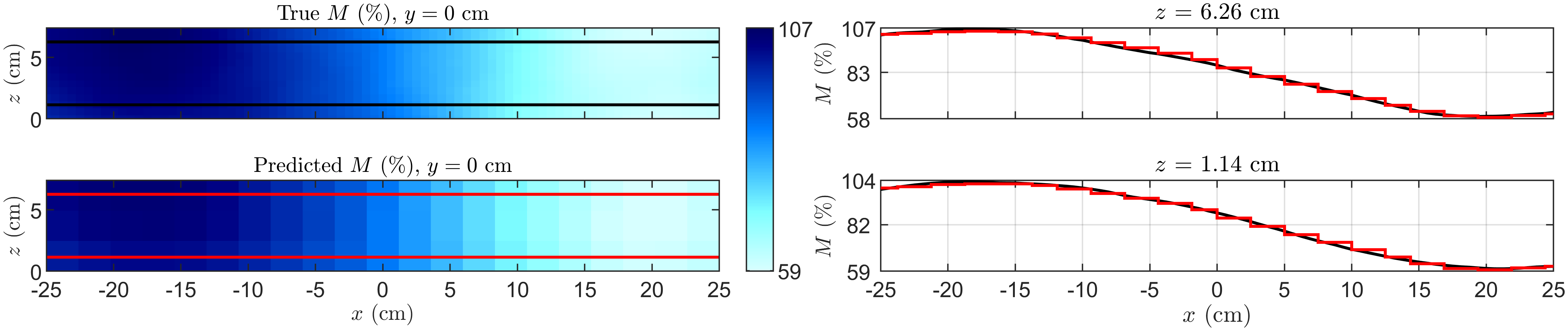}
\caption{\label{fig:results} Three example cases having low (top), moderate (middle), and high (bottom) moisture content. Pictures on the left show the true and predicted field values and pictures on the right for two fixed $z$ values, respectively. The selected lines are visualized with black and red horizontal lines on the corresponding field graphs.}
\end{figure*}

Three test samples with low, moderate, and high moisture contents are chosen and the corresponding scattered fields are measured and given as an input to the trained neural network. The noise level is set to $\beta=0.005$, see Eq. (\ref{eq:noisemodel}), which leads to ${\rm SNR_{dB}} = 28.7\, {\rm dB}$. The true test samples and predicted outputs from the neural network are shown in Fig. \ref{fig:results}. Further to assess the closeness of the estimate to the true case, moisture values on two horizontal lines are visualised and shown in the right column of Fig. \ref{fig:results}. For all selected samples, the network provides very good estimation accuracy. 

\subsection{Sample with nearly uniform distribution}\label{sec:uni}

\begin{figure*}[!htb]
\centering
\includegraphics[width=0.99\textwidth]{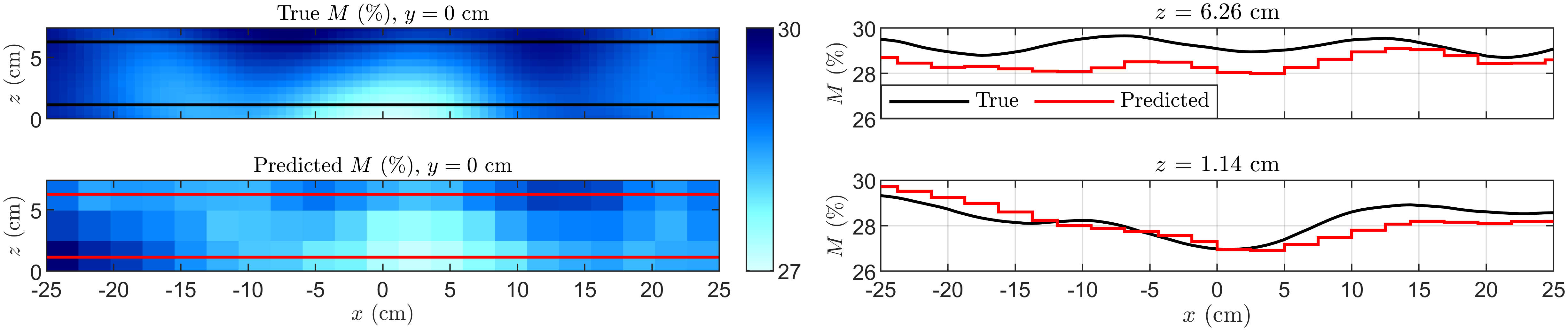}
\includegraphics[width=0.99\textwidth]{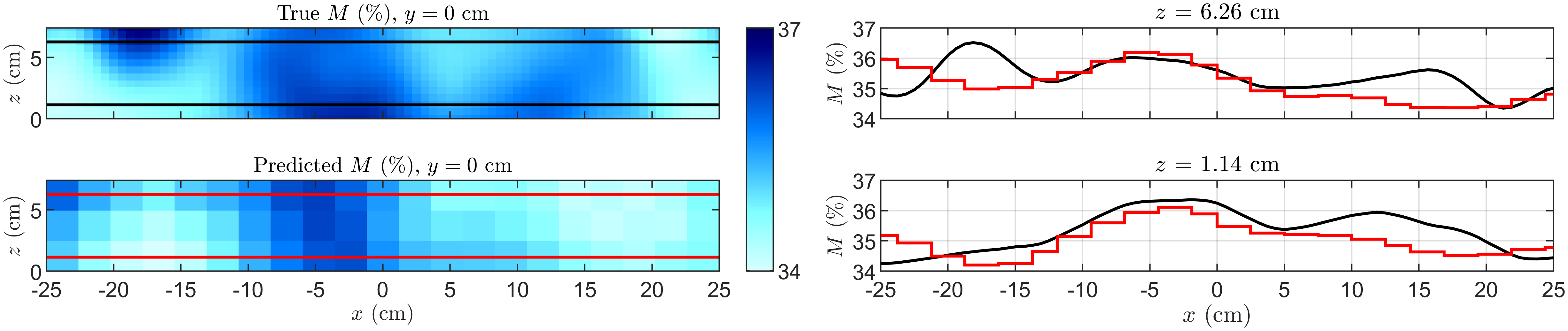}
\caption{\label{fig:results_const} Two example cases having minor variations within the moisture content distribution. Otherwise the same caption as in Fig. \ref{fig:results}.}
\end{figure*}

In the actual drying process, the aim is that the moisture content distribution of the foam is nearly uniform when it comes out from the drying system (see Fig. \ref{fig:scemu}). Therefore, the actual training and also the test data can contain samples, where the moisture content variations are minor (see Eq. (\ref{eq:moist})). Two example cases, in which the criterion $\max(M)-\min(M) < 3 \%$ is fulfilled, are shown in Fig. \ref{fig:results_const}. As in Section \ref{sec:low}, the noise parameter $\beta$ in (\ref{eq:noisemodel}) is set to 0.005 (${\rm SNR_{dB}} = 28.7\, {\rm dB}$). Estimation accuracy is good for both cases.

\subsection{Generalization capabilities of the network}\label{sec:gen}

\subsubsection{Inclusions}

\begin{figure*}[!htb]
\centering
\includegraphics[width=0.99\textwidth]{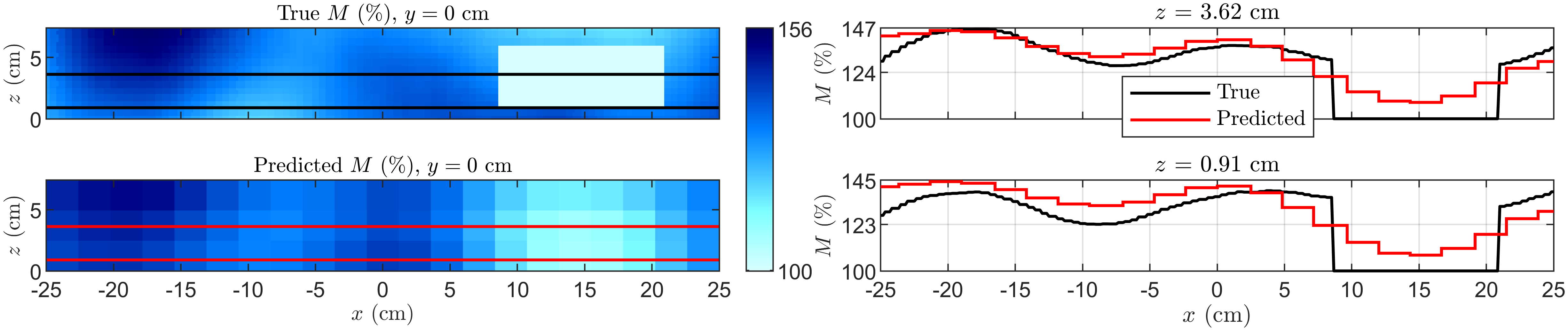}
\includegraphics[width=0.99\textwidth]{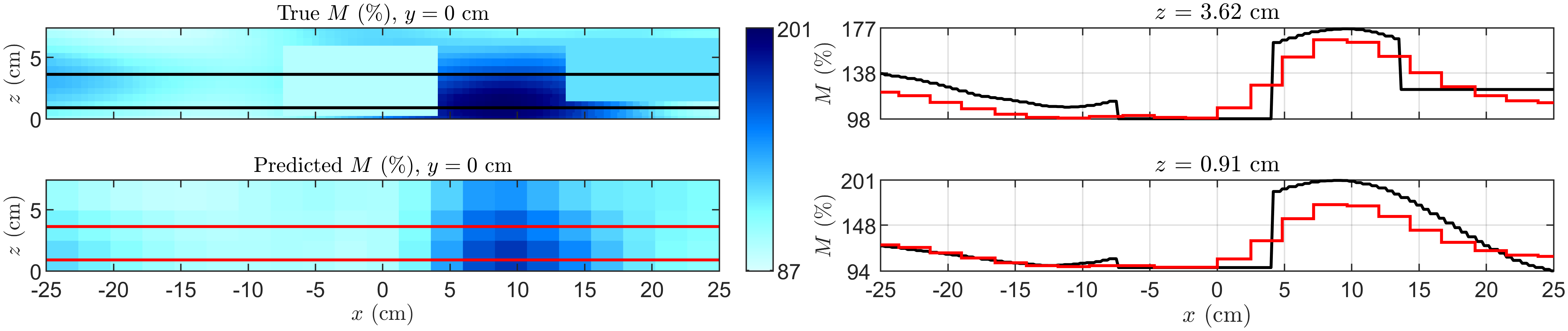}
\caption{\label{fig:results_extr} Two example cases having rectangular cuboid shaped discontinuities (top: one inclusion, bottom: two inclusions) within the moisture content distribution. Otherwise the same caption as in Fig. \ref{fig:results}.}
\end{figure*}

The test samples with one cuboid inclusion discontinuity in the first case and two cuboid inclusions discontinuity in the second case are chosen. The noise level is assumed to be low ($\beta = 0.005,\ {\rm SNR_{dB}} = 28.7\, {\rm dB}$). The scattered fields for the two cases are simulated and given as input to the trained neural network. The true moisture content distributions and predicted outputs are shown in Fig. \ref{fig:results_extr}. The estimation accuracy is fairly good for both test cases. 

\subsubsection{Very high moisture content}\label{sec:hspot}

\begin{figure*}[!htb]
\centering
\includegraphics[width=0.99\textwidth]{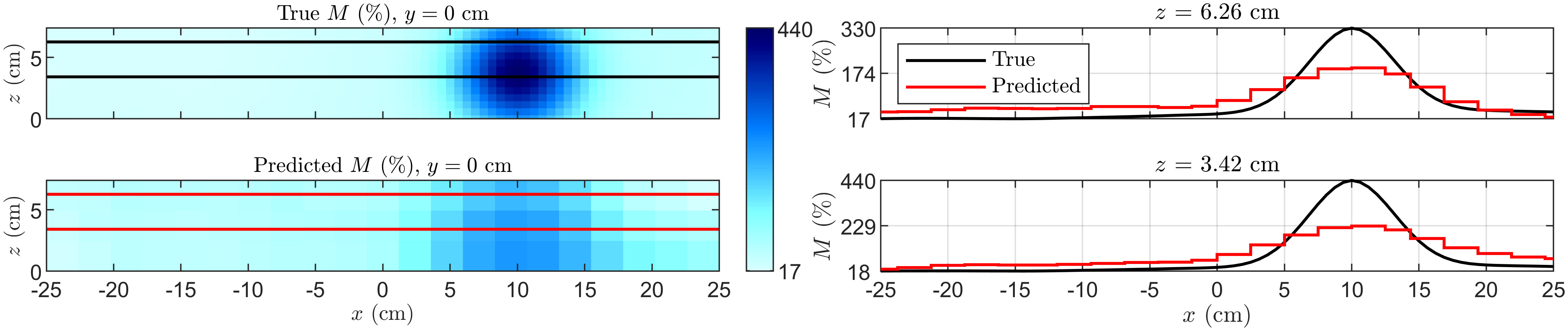}
\caption{\label{fig:results_extra} An example case having a moisture content distribution that has an additional component that adds a strong positive peak to the true moisture content distribution. As in Fig. \ref{fig:results}, pictures on the left show the true and predicted field values and pictures on the right for two rows. The selected lines are visualized with black and red horizontal lines on the corresponding field graphs.}
\end{figure*}

In this experiment, we want to demonstrate the network's capabilities to predict larger moisture values than found in the training database, by adding a very high moisture level spot to the true moisture field (see Fig. \ref{fig:results_extra} (topleft)) while the noise level is assumed to be low ($\beta = 0.005,\ {\rm SNR_{dB}} = 28.7\, {\rm dB}$). The estimation result shown in Fig. \ref{fig:results_extra} illustrates that the network is not capable of recovering the true distribution. We observe that the shape of the estimate is roughly the same as it is in the true distribution but the amplitude information is wrong. Result suggests that the network fails to predict higher values than found in the training data.

\subsection{Robustness to measurement noise} \label{sec:noise}

\begin{figure*}[!htb]
\centering
\includegraphics[width=0.99\textwidth]{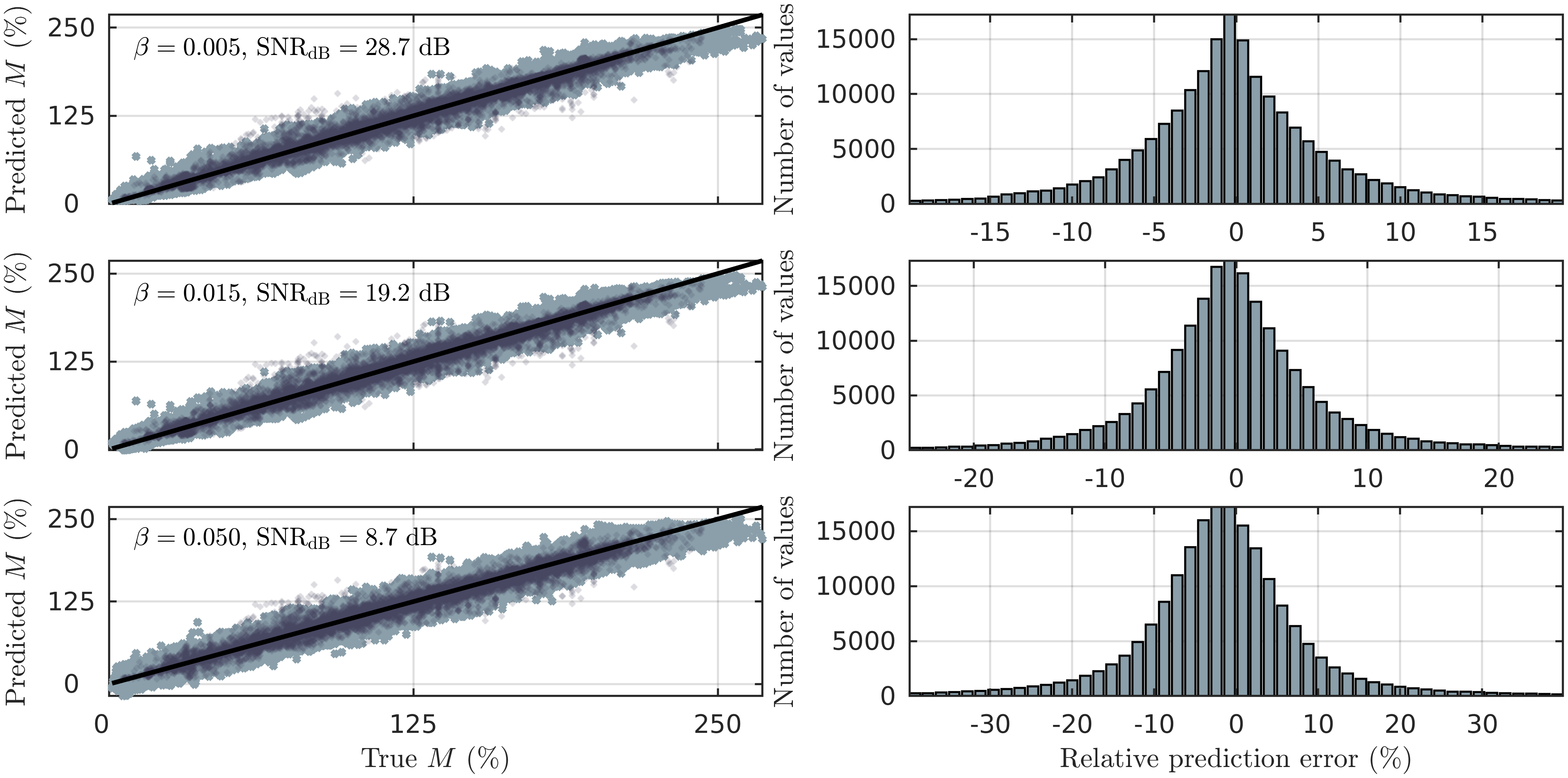}
\caption{\label{fig:results_n1}   Predicted pixel values for moisture density  for the test datasets (light gray dots corresponds to the dataset without rectangular cuboid shaped inclusions) with different values of $\beta$ (see Eq. (\ref{eq:noisemodel})). Figures on the right show histograms of the differences between the predicted and true values (relative prediction error).}
\end{figure*}

We wanted to study how sensitive the network is to noise and hence evaluated the network with different levels of noise. Estimates for the both test datasets, contaminated with the white noise component of 
\begin{enumerate}
    \item a low ($\beta=0.005,\ {\rm SNR_{dB}} = 28.7\, {\rm dB}$) and high level ($\beta=0.015,\ {\rm SNR_{dB}} = 19.2\, {\rm dB}$) when compared to the noise level assumed in the training ($\beta\in [0.001,\ 0.022]$), and
    \item higher ($\beta=0.05,\ {\rm SNR_{dB}} = 8.7\, {\rm dB}$) than the assumed maximum in the training
\end{enumerate}
are shown in Fig. \ref{fig:results_n1}, respectively. Results for different datasets are drawn in different color. The figures also include relative prediction error histograms.

\section{Discussion}\label{sec:Discussion}
To study the estimation accuracy, a synthetic test dataset containing 3000 samples was generated. The network recovers the cross-sectional moisture content distribution inside the porous foam with acceptable accuracy, as shown in Fig. \ref{fig:results_n1} (top and middle). To illustrate the network operation in sample level, estimates for five moisture content distribution scenarios are shown in Figs. \ref{fig:results} and \ref{fig:results_const}. Three of the selected cases contain non-uniform low, moderate, and high moisture content distributions while in two cases the distribution is almost uniform. In all cases, the network performs very well.

The network architecture has an impact on estimation accuracy. We tested different network architectures by varying the number of layers and the number of neurons in each layer.  For the convolution layers, we tested how the filter size affects the network's behavior. In our model, the filter size remained constant between the layers of the network. In addition to network architecture, we tested the effect of different learning rate values for the selected Adam optimizer and the batch size. From the tested configurations, the selected was the best performing. A similar type of network architecture has proven to be effective also in other wave-based inverse problems studied by the authors \cite{lahivaara18, lahivaara19}.

After fixing the network architecture, two main components are affecting the accuracy of the predictions. Firstly, the uncertainties related to the (exponential) model (\ref{eq:param}) that maps the moisture content value to physical parameters $\epsilon_r^\prime$ and $\epsilon_r^{\prime\prime}$ have an impact on the network performance. Uncertainties, listed in Table \ref{tab:parameters}, causes that especially for low $M$ values, the model gives quite wide range of values for corresponding physical parameters $\epsilon_r^\prime$ and $\epsilon_r^{\prime\prime}$. Therefore, as precise as possible measurements are needed to get an accurate model (\ref{eq:param}). It can be expected that the relative error distributions, shown in Fig. \ref{fig:results_n1} (topright and middleright), are narrower with less uncertain physical mapping model.

Secondly, the scattering, with current problem setup, can be expected to have a minor effect on the recorded signals for low moisture content values since the physical parameters $\epsilon_r^\prime$ and $\epsilon_r^{\prime\prime}$ are very close to air values. Similarly, the low moisture content variation inside the foam has a minor effect on the wave scattering with selected operating parameters. If the noise level is higher than the scattering effect seen in the recorded signals, the information of the moisture content distribution is lost and the estimation accuracy is expected to be poor.

 We also studied the network's generalization capabilities. We generated a test dataset which comprises of 100 samples. These samples contain fictitious rectangular cuboid  shaped inclusions inside the foam. The estimation accuracy remained in a similar level as for the original test data containing smooth variations in the moisture field, as shown in Fig. \ref{fig:results_n1}. One must note that here the moisture content amplitudes are inside the amplitude values found in the training database which (at least partially) enables the network to predict. On the other hand, as discussed already in Section \ref{sec:hspot} and shown in Fig. \ref{fig:results_extra}, the network fails to predict the moisture content distribution which has values that are not covered by the actual training data. The result shows that the network recovers the distribution shape with acceptable accuracy but fails in the amplitude estimation. Also, the estimation accuracy decreases when the noise amplitude is beyond the training data (see Fig. \ref{fig:results_n1} (bottom)). In this latter case, the estimation accuracy is also potentially affected by the fact that the information about the moisture content variations is swamped with the noise.

All the estimation results presented in this work show a piece-wise spatial variation of the moisture distribution in comparison to the smooth variation of the moisture in the respective true cases. The piece-wise variation in the estimation results is caused due to less number of pixels in the output image. Thus, to achieve smoothness in the estimation results the total number of pixels in the output needs to be increased. However, the drying process itself, which is based on microwave processing in a cavity at $2.45\ {\rm GHz}$, may allow control of moisture distribution with a spatial resolution of about half wavelength at maximum. Therefore, the knowledge of moisture distribution with a spatial resolution significantly better than half wavelength, will not necessarily result in better process control.

In all test cases, the trained neural network estimated the moisture content in the foam in less than a second. Therefore, the neural network only needs to be trained in offline mode and can then be applied online to get real-time estimates. For our 3D estimation problem, it is a significant improvement in comparison to commonly used gradient-based iterative methods. These methods are often coupled with line search and require the calculation of gradient (at each iteration) both of which increase the computational time. With the system configuration explained in Section \ref{sec:MWT}, the computation of 220 $S$-parameter values (one complete forward problem solution) takes approximately 10 minutes. Hence, the overall computational time needed with a common gradient-based iterative inverse problem solver, even with a small number of iterations, becomes unbearable.

\section{Conclusion}\label{sec:Conc}

In this paper, a neural network coupled with MWT is shown to be numerically feasible to be integrated with a microwave drying system. The estimation results are obtained in milliseconds, which shows the promising capability of the present method to be used in real-time estimation. However, the underlying uncertainties in the parametric model, which relates the moisture content and temperature values to dielectric values, can bring down the performance of this study when applied to actual measurements. Therefore, dielectric measurements from different methods are under investigation at a laboratory scale to define a more accurate parametric model. In this feasibility study, the antenna type is not considered and the estimation results are shown for simple antenna type as the main focus is to evaluate the performance of the neural network-based approach. Therefore, for a demonstration with actual data, the designed network will be re-trained with a more reliable parametric model and suitable antenna type. Here, the studied microwave imaging modality is applied to recover moisture content distribution inside a porous foam but the framework is applicable to investigate other material types together with different physical parameters.

\section*{Aknowledgements}

This work has been supported by the Academy of Finland (Finnish Centre
of Excellence of Inverse Modelling and Imaging, project number 312344,
and project number 321761) and by the European Union's Horizon 2020
Research and Innovation Programme under the Marie Sk\l{}odowska-Curie
grant agreement No. 764902 (TOMOCON - www.tomocon.eu).


\begin{thebibliography}{10}

\bibitem{comsol}
{COMSOL} {M}ultiphysics$\textsuperscript{TM}$ v. 5.3a, 2020.
\newblock www.comsol.com, {COMSOL} {AB}, Stockholm, Sweden.

\bibitem{tensorflow2015}
M.~Abadi, A.~Agarwal, P.~Barham, E.~Brevdo, Z.~Chen, C.~Citro, G.~S. Corrado,
  A.~Davis, J.~Dean, M.~Devin, S.~Ghemawat, I.~Goodfellow, A.~Harp, G.~Irving,
  M.~Isard, Y.~Jia, R.~Jozefowicz, L.~Kaiser, M.~Kudlur, J.~Levenberg,
  D.~Man\'{e}, R.~Monga, S.~Moore, D.~Murray, C.~Olah, M.~Schuster, J.~Shlens,
  B.~Steiner, I.~Sutskever, K.~Talwar, P.~Tucker, V.~Vanhoucke, V.~Vasudevan,
  F.~Vi\'{e}gas, O.~Vinyals, P.~Warden, M.~Wattenberg, M.~Wicke, Y.~Yu, and
  X.~Zheng.
\newblock {TensorFlow}: Large-scale machine learning on heterogeneous systems,
  2015.
\newblock Software available from tensorflow.org.

\bibitem{10}
A.~Abubakar, P.~M. van~den Berg, and S.~Y. Semenov.
\newblock Two- and three-dimensional algorithms for microwave imaging and
  inverse scattering.
\newblock {\em Journal of Electromagnetic Waves and Applications},
  17(2):209--231, 2003.

\bibitem{728803}
P.~G. Bartley, S.~O. Nelson, R.~W. McClendon, and S.~Trabelsi.
\newblock Determining moisture content of wheat with an artificial neural
  network from microwave transmission measurements.
\newblock {\em IEEE Transactions on Instrumentation and Measurement},
  47(1):123--126, 1998.

\bibitem{7811280}
A.~S. Beaverstone, D.~S. Shumakov, and N.~K. Nikolova.
\newblock Frequency-domain integral equations of scattering for complex scalar
  responses.
\newblock {\em IEEE Transactions on Microwave Theory and Techniques},
  65(4):1120--1132, 2017.

\bibitem{1202957}
E.~Bermani, A.~Boni, S.~Caorsi, and A.~Massa.
\newblock An innovative real-time technique for buried object detection.
\newblock {\em IEEE Transactions on Geoscience and Remote Sensing},
  41(4):927--931, 2003.

\bibitem{buduma17}
N.~Buduma.
\newblock {\em Fundamentals of Deep Learning, Designing Next-Generation Machine
  Intelligence Algorithms}.
\newblock O'Reilly Media, 2017.

\bibitem{752198}
S.~Caorsi and P.~Gamba.
\newblock Electromagnetic detection of dielectric cylinders by a neural network
  approach.
\newblock {\em IEEE Transactions on Geoscience and Remote Sensing},
  37(2):820--827, 1999.

\bibitem{5271011}
W.~C. Chew.
\newblock {\em Waves and Fields in Inhomogenous Media}.
\newblock IEEE Press, 1995.

\bibitem{312725}
I.~Elshafiey, L.~Udpa, and S.~S. Udpa.
\newblock Application of neural networks to inverse problems in
  electromagnetics.
\newblock {\em IEEE Transactions on Magnetics}, 30(5):3629--3632, 1994.

\bibitem{560338}
A.~Franchois and C.~Pichot.
\newblock Microwave imaging-complex permittivity reconstruction with a
  {L}evenberg-{M}arquardt method.
\newblock {\em IEEE Transactions on Antennas and Propagation}, 45(2):203--215,
  1997.

\bibitem{5617228}
M.~Haynes and M.~Moghaddam.
\newblock Multipole and s-parameter antenna and propagation model.
\newblock {\em IEEE Transactions on Antennas and Propagation}, 59(1):225--235,
  2011.

\bibitem{7496964}
N.~Javed, A.~Habib, Y.~Amin, J.~Loo, A.~Akram, and H.~Tenhunen.
\newblock Directly printable moisture sensor tag for intelligent packaging.
\newblock {\em IEEE Sensors Journal}, 16(16):6147--6148, 2016.

\bibitem{kaipio07}
J.~Kaipio and E.~Somersalo.
\newblock Statistical inverse problems: {D}iscretization, model reduction and
  inverse crimes.
\newblock {\em Journal of Computational and Applied Mathematics},
  198(2):493--504, 2007.

\bibitem{Kingma2014}
D.~P. {Kingma} and J.~Ba.
\newblock {Adam: A Method for Stochastic Optimization}.
\newblock {\em ArXiv e-prints}, 2014.

\bibitem{krizhevsky}
A.~Krizhevsky, I.~Sutskever, and G.~Hinton.
\newblock Imagenet classification with deep convolutional neural networks.
\newblock In F.~Pereira, C.~J.~C. Burges, L.~Bottou, and K.~Q. Weinberger,
  editors, {\em Advances in Neural Information Processing Systems 25}, pages
  1097--1105. Curran Associates, Inc., 2012.

\bibitem{lahivaara18}
T.~L\"ahivaara, L.~K\"arkk\"ainen, J.~M. Huttunen, and J.~S. Hesthaven.
\newblock Deep convolutional neural networks for estimating porous material
  parameters with ultrasound tomography.
\newblock {\em The Journal of the Acoustical Society of America},
  143(2):1148--1158, 2018.

\bibitem{lahivaara19}
T.~L\"ahivaara, A.~Malehmir, L.~K\"arkk\"ainen, J.~M. Huttunen, and J.~S.
  Hesthaven.
\newblock Estimation of groundwater storage from seismic data using deep
  learning.
\newblock {\em Geophysical Prospecting}, 67(8):2115--2126, 2019.

\bibitem{8565987}
L.~{Li}, L.~G. {Wang}, F.~L. {Teixeira}, C.~{Liu}, A.~{Nehorai}, and T.~J.
  {Cui}.
\newblock Deepnis: Deep neural network for nonlinear electromagnetic inverse
  scattering.
\newblock {\em IEEE Transactions on Antennas and Propagation},
  67(3):1819--1825, 2019.

\bibitem{LIM2003159}
M.~Lim, K.~Lim, and M.~Abdullah.
\newblock Rice moisture imaging using electromagnetic measurement technique.
\newblock {\em Food and Bioproducts Processing}, 81(3):159 -- 169, 2003.

\bibitem{guido}
G.~Link and V.~Ramopoulos.
\newblock Simple analytical approach for industrial microwave applicator
  design.
\newblock {\em Chemical Engineering and Processing - Process Intensification},
  125:334--342, 2018.

\bibitem{24}
Z.~Meng, Z.~Wu, and J.~Gray.
\newblock Microwave sensor technologies for food evaluation and analysis:
  Methods, challenges and solutions.
\newblock {\em Transactions of the Institute of Measurement and Control},
  40(12):3433--3448, 2018.

\bibitem{7801816}
Z.~Meng, Z.~Wu, C.~Muvianto, and J.~Gray.
\newblock A data-oriented m2m messaging mechanism for industrial iot
  applications.
\newblock {\em IEEE Internet of Things Journal}, 4(1):236--246, 2017.

\bibitem{monk03}
P.~Monk.
\newblock {\em Finite Element Methods for Maxwell's Equations}.
\newblock Oxford University Press, 2003.

\bibitem{Okamura2000}
S.~Okamura.
\newblock Microwave technology for moisture measurement.
\newblock {\em Subsurface Sensing Technologies and Applications},
  1(2):205--227, 2000.

\bibitem{6008622}
M.~Ostadrahimi, P.~Mojabi, C.~Gilmore, A.~Zakaria, S.~Noghanian, S.~Pistorius,
  and J.~LoVetri.
\newblock Analysis of incident field modeling and incident/scattered field
  calibration techniques in microwave tomography.
\newblock {\em IEEE Antennas and Wireless Propagation Letters}, 10:900--903,
  2011.

\bibitem{rasmussen}
C.~Rasmussen and C.~Williams.
\newblock {\em Gaussian Processes for Machine Learning}.
\newblock The MIT Press, 2006.

\bibitem{Makul}
P.~Rattanadecho and N.~Makul.
\newblock Microwave-assisted drying: A review of the state-of-the-art.
\newblock {\em Drying Technology}, 34(1):1--38, 2016.

\bibitem{8709721}
Y.~{Sanghvi}, Y.~N. G.~B. {Kalepu}, and U.~{Khankhoje}.
\newblock Embedding deep learning in inverse scattering problems.
\newblock {\em IEEE Transactions on Computational Imaging}, 2020.
\newblock in press.

\bibitem{Scherer}
G.~W. Scherer.
\newblock Theory of drying.
\newblock {\em Journal of the American Ceramic Society}, 73(1):3--14, 1990.

\bibitem{7951691}
B.~L. Shrestha, H.~C. Wood, L.~Tabil, O.~Baik, and S.~Sokhansanj.
\newblock Microwave permittivity-assisted artificial neural networks for
  determining moisture content of chopped alfalfa forage.
\newblock {\em IEEE Instrumentation Measurement Magazine}, 20(3):37--42, 2017.

\bibitem{sun16}
Y.~Sun.
\newblock {\em Adaptive and Intelligent Temperature Control of Microwave
  Heating Systems with Multiple Sources}.
\newblock PhD thesis, KIT Scientific Publishing, Karlsruhe, 2016.

\bibitem{Sun:18}
Y.~Sun, Z.~Xia, and U.~S. Kamilov.
\newblock Efficient and accurate inversion of multiple scattering with deep
  learning.
\newblock {\em Opt. Express}, 26(11):14678--14688, 2018.

\bibitem{8094000}
S.~Trabelsi.
\newblock New calibration algorithms for dielectric-based microwave moisture
  sensors.
\newblock {\em IEEE Sensors Letters}, 1(6):1--4, 2017.

\bibitem{Samir}
S.~Trabelsi, A.~W. Kraszewski, and S.~O. Nelson.
\newblock A microwave method for on-line determination of bulk density and
  moisture content of particulate materials.
\newblock {\em IEEE Trans. Instrumentation and Measurement}, 47:127--132, 1998.

\bibitem{1634890}
S.~Trabelsi and S.~O. Nelson.
\newblock Nondestructive sensing of physical properties of granular materials
  by microwave permittivity measurement.
\newblock {\em IEEE Transactions on Instrumentation and Measurement},
  55(3):953--963, 2006.

\bibitem{Trabelsi2009}
S.~Trabelsi, S.~O. Nelson, and M.~A. Lewis.
\newblock Microwave nondestructive sensing of moisture content in shelled
  peanuts independent of bulk density and with temperature compensation.
\newblock {\em Sensing and Instrumentation for Food Quality and Safety},
  3(2):114--121, 2009.

\bibitem{5265547}
J.~L. Volakis, A.~Chatterjee, and L.~C. Kempel.
\newblock {\em Finite Element Method Electromagnetics: Antennas, Microwave
  Circuits, and Scattering Applications}.
\newblock IEEE Press, 1998.

\bibitem{25}
F.~V\"olgyi.
\newblock Monitoring of particleboard production using microwave sensors.
\newblock {\em Sensors Update}, 7(1):249--274, 2000.

\bibitem{17}
Y.~Wang and X.~Gong.
\newblock A neural network approach to microwave imaging.
\newblock {\em International Journal of Imaging Systems and Technology},
  11(3):159--163, 2000.

\bibitem{8476623}
Z.~Wei and X.~Chen.
\newblock Deep-learning schemes for full-wave nonlinear inverse scattering
  problems.
\newblock {\em IEEE Transactions on Geoscience and Remote Sensing}, pages
  1--12, 2018.

\bibitem{8741152}
Z.~{Wei} and X.~{Chen}.
\newblock Physics-inspired convolutional neural network for solving full-wave
  inverse scattering problems.
\newblock {\em IEEE Transactions on Antennas and Propagation},
  67(9):6138--6148, 2019.

\bibitem{15}
Z.~Wu.
\newblock Developing a microwave tomographic system for multiphase flow
  imaging: advances and challenges.
\newblock {\em Transactions of the Institute of Measurement and Control},
  37(6):760--768, 2015.

\bibitem{8016570}
Z.~Wu and H.~Wang.
\newblock Microwave tomography for industrial process imaging: Example
  applications and experimental results.
\newblock {\em IEEE Antennas and Propagation Magazine}, 59(5):61--71, 2017.

\bibitem{6482167}
K.~Y. You, C.~Y. Lee, Y.~L. Then, S.~H.~C. Chong, L.~L. You, Z.~Abbas, and
  E.~M. Cheng.
\newblock Precise moisture monitoring for various soil types using handheld
  microwave-sensor meter.
\newblock {\em IEEE Sensors Journal}, 13(7):2563--2570, 2013.

\bibitem{7422005}
Y.~Zhong, M.~Lambert, D.~Lesselier, and X.~Chen.
\newblock A new integral equation method to solve highly nonlinear inverse
  scattering problems.
\newblock {\em IEEE Transactions on Antennas and Propagation},
  64(5):1788--1799, 2016.

\end{thebibliography}

\end{document}